\documentclass{PoS}
\usepackage{epsfig}

\PoS{PoS(LAT2005)199}

\title{ 
Finite-temperature chiral transition in QCD with quarks in the
fundamental and adjoint representation.
} 

\ShortTitle{Finite-temperature chiral transition in QCD with quarks 
in the fundamental and adjoint representation }

\author{Francesco Basile\\ Dipartimento di
Fisica dell'Universit\`a di Pisa and INFN, \\ Via Buonarroti 2, 56127 Pisa,
Italy. 
E-mail: \email{basile@sns.it} } 
\author{Andrea Pelissetto\\
Dipartimento di Fisica dell'Universit\`a di Roma ``La Sapienza'' and
INFN, \\ Piazzale Moro 2, 00185 Roma, Italy.
E-mail: \email{Andrea.Pelissetto@roma1.infn.it} } 
\author{\speaker{Ettore Vicari} \\
Dipartimento di Fisica dell'Universit\`a di Pisa and INFN, \\ Via
Buonarroti 2, 56127 Pisa, Italy.
E-mail: \email{vicari@df.unipi.it} 
}

\abstract{ We study the nature of the finite-temperature chiral
transition in QCD with $N_f$ light quarks in the fundamental and
adjoint representation.  Universality and renormalization-group (RG)
arguments show that the possibility of having a continuous transition
is related to the existence of a stable fixed point (FP) in the RG
flow of a 3D Landau-Ginzburg-Wilson $\Phi^4$ theory with the same
chiral symmetry-breaking pattern. The RG flow of these theories is
studied by field-theoretical approaches, computing and analyzing
high-order perturbative series, up to six loops.  According to this RG
analysis, the transition in QCD can be continuous only for $N_f=2$.
In this case it belongs to the 3D O(4) universality class.  We also
find a stable FP corresponding to a 3D universality class with
symmetry breaking U(2)$_L$$\otimes$U(2)$_R$$\rightarrow$U(2)$_V$,
which implies that the transition can be continuous also if
the axial-anomaly effects are suppressed at $T_c$.  In the case
of quarks in the adjoint representation, we can have a continuous
transition for $N_f=1,2$.  For $N_f=1$ it belongs to the O(3)
universality class.  For $N_f=2$ it belongs to a new 3D universality
class characterized by the symmetry breaking SU(4)$\rightarrow$SO(4).
}

\FullConference{XXIIIrd International Symposium on Lattice Field Theory\\
25-30 July 2005\\
Trinity College, Dublin, Ireland}

\begin{document}

The thermodynamics of Quantum Chromodynamics (QCD) is characterized by
a transition from a low-$T$ hadronic phase, in which chiral symmetry
is broken, to a high-$T$ phase with deconfined quarks and gluons
(quark-gluon plasma), in which chiral symmetry is restored
\cite{reviews}.  Our understanding of the finite-$T$ phase transition
is essentially based on the relevant symmetry and symmetry-breaking
pattern (SBP). In the presence of $N_f$ light quarks the relevant
symmetry is the chiral symmetry ${\rm U}(1)_V\otimes{\rm
SU}(N_f)_L\otimes{\rm SU}(N_f)_R $. At $T=0$ this symmetry is
spontaneously broken to U(1)$_V\otimes$SU($N_f$)$_V$ with a nonzero
quark condensate $\langle {\bar \psi} \psi \rangle$.  The finite-$T$
transition is related to the restoring of
the chiral symmetry.  It is therefore characterized by an $N_f\times
N_f$ complex-matrix order parameter $\Phi_{ij}$, related to the
bilinear quark operator $\bar{\psi}_{Li}\psi_{Rj}$, and the SBP
\begin{equation}
{\rm SU}(N_f)_L\otimes {\rm SU}(N_f)_R   \rightarrow {\rm SU}(N_f)_V . 
\label{qcdsb}
\end{equation}
If the axial U(1)$_A$
symmetry is effectively restored at $T_c$, the expected SBP becomes
\begin{eqnarray}
{\rm U}(N_f)_L\otimes {\rm U}(N_f)_R  \rightarrow {\rm U}(N_f)_V .
\label{qcdsbwa}
\end{eqnarray}
Lattice Monte Carlo (MC) simulations suggest that these is not the
case in three-color QCD. However, since the anomaly gets suppressed in
the large-$N_c$ limit ($\partial_\mu J_5^\mu \propto {1\over N_c} Q$),
SBP (\ref{qcdsbwa}) may be relevant in the large-$N_c$ limit.

Although deconfinement and chiral symmetry restoration are apparently
related to different mechanisms, they seem to be somehow coupled in
QCD, since lattice computations show that the Polyakov loop has a
sharp increase at $T_c$ where the chiral condensate vanishes.
However, the interplay between the two effects is not clear yet.
Insight into this question may be gained by investigating related
models, such as ${\rm SU}(N_c)$ gauge theories with $N_f$ Dirac
fermions in the adjoint representation (aQCD).\footnote{AQCD is 
asymptotically free only for $N_f < 11/4$, thus only
the cases $N_f=1,2$ are interesting.}  Unlike QCD, aQCD is also
invariant under global ${\mathbb Z}_{N_c}$ transformations related to
the center of the gauge group SU($N_c$), as in pure ${\rm SU}(N_c)$
gauge theories. There are two well-defined order parameters in the
light-quark regime, related to the confining and chiral modes,
i.e. the Polyakov loop and the quark condensate. One generally expects
two transitions: a deconfinement transition at $T_d$ associated with
the breaking of the ${\mathbb Z}_{N_c}$ symmetry, and a chiral
transition at $T_c$ in which chiral symmetry is restored. In aQCD with
$N_f$ massless flavors the chiral-symmetry group extends to ${\rm
SU}(2N_f)$. At $T=0$ this symmetry is expected to spontaneously break
to ${\rm SO}(2N_f)$, due to quark condensation.  Therefore the
SBP at the finite-$T$ chiral transition is expected to be
\begin{equation}
{\rm SU}(2N_f) \rightarrow {\rm SO}(2N_f) 
\label{aqcdsb}
\end{equation}
with a  symmetric $2N_f\times 2N_f$ complex matrix as order parameter
related to the bilinear quark condensate.  If the axial U(1)$_A$
symmetry is restored at $T_c$, the SBP is
\begin{eqnarray}
{\rm U}(2N_f) \rightarrow {\rm O}(2N_f) .
\label{aqcdsbwa}
\end{eqnarray}
MC simulations for $N_c=3$ and $N_f=2$ \cite{KL-99,EHS-05} show that
the deconfinement transition at $T_d$ is first order.  Results at the
chiral transition appear consistent with a continuous transition.
Interestingly, the ratio between the two critical temperatures turns
out to be quite large, $T_c/T_d\approx 8$, suggesting a rather weak
interplay between the corresponding underlying mechanisms.

In order to study the nature of the finite-$T$ chiral transition in
QCD and aQCD, we exploit universality and
renormalization-group (RG) arguments, as originally applied by
Pisarski and Wilczek in Ref.~\cite{PW-84}.  They can be summarized as
follows.

\medskip

\noindent (i)
Let us first assume that the phase transition at $T_c$ is continuous
for vanishing quark masses.  In this case the length scale of the
critical modes diverges approaching $T_c$, becoming eventually much
larger than $1/T_c$, which is the size of the euclidean ``temporal''
dimension at $T_c$.  Therefore, the asymptotic critical behavior must
be associated with a 3D universality class with the same SBP. 

\noindent (ii) The existence of such a 3D universality class can be
investigated by considering the most general Landau-Ginzburg-Wilson
(LGW) $\Phi^4$ theory compatible with the given SBP, which
describes the critical modes at $T_c$.  Neglecting the U(1)$_A$ anomaly,
it is given by
\begin{eqnarray}
{\cal L}_{{\rm U}(N)} = {\rm Tr} (\partial_\mu \Phi^\dagger) (\partial_\mu \Phi)
+r {\rm Tr} \Phi^\dagger \Phi 
+ {u_0\over 4} \left( {\rm Tr} \Phi^\dagger \Phi \right)^2
+ {v_0\over 4} {\rm Tr} \left( \Phi^\dagger \Phi \right)^2 .
\label{LUN}
\end{eqnarray}
If $\Phi_{ij}$ is a generic $N\times N$ complex matrix, the symmetry
is U$(N)_L\otimes$U$(N)_R$, which breaks to U$(N)_V$ if $v_0>0$, thus
providing the LGW theory relevant for QCD with $N_f=N$.  If
$\Phi_{ij}$ is also symmetric, the global symmetry is U$(N)$, which
breaks to O($N$) if $v_0>0$, which is the case relevant for aQCD with
$N_f=N/2$.  The reduction of the symmetry to
SU$(N_f)_L\otimes$SU$(N_f)_R$ for QCD [SU$(2N_f)$ for aQCD], due to
the axial anomaly, is achieved by adding determinant terms, such as
\begin{eqnarray}
{\cal L}_{{\rm SU}(N)} =  {\cal L}_{{\rm  U}(N)} +
w_0 \left( {\rm det} \Phi^\dagger + {\rm det} \Phi \right) .
\label{LSUN}
\end{eqnarray}
Nonvanishing quark masses can be accounted for by adding 
an external-field term ${\rm Tr} \left( H \Phi + {\rm h.c.}\right)$.

\noindent (iii) The critical behavior at a continuous transition is
determined by the fixed points (FPs) of the RG flow: the absence of a
stable FP generally implies first-order transitions.  Therefore, a
necessary condition of consistency with the initial hypothesis (i) of
a continuous transition is the existence of stable FP in the
corresponding LGW $\Phi^4$ theory.  If no stable FP exists, the finite-$T$
chiral transition of QCD (aQCD) is predicted to be first order.  If a
stable FP exists, the transition can be continuous, and its universal
critical behavior is determined by the FP; but this does not exclude a
first-order transition if the system is outside the attraction domain
of the stable FP.

\medskip

The above arguments show that the nature of the finite-$T$ transition
in QCD and aQCD can be investigated by studying the RG flow of the
corresponding 3D LGW $\Phi^4$ theories. For this purpose we consider
two different field-theoretical (FT) perturbative approaches.  One is
defined within the massive (disordered) phase using a zero-momentum
renormalization (MZM) scheme.  The other one is defined within the
massless (critical) theory using a minimal subtraction ($\overline{\rm
MS}$) scheme (we consider a 3D-$\overline{\rm MS}$ scheme without
$\epsilon$ expansion). The RG flow is determined by the FPs, which are
given by the common zeroes of the $\beta$-functions associated with
the quartic couplings.  A FP is stable
if all eigenvalues of its stability matrix have positive real
part. Using symbolic manipulations programs, we computed the expansion
up to six loops in the MZM scheme (which requires the calculation of
approximately 1500 Feynman diagrams) and up to five loops in the
$\overline{\rm MS}$ scheme. Since perturbative FT expansions are
asymptotic, it is necessary to resum the series. This is done by
exploiting Borel summability and knowledge of the large-order
behavior, which is inferred by semiclassical calculations of instanton
solutions.  We first resum the $\beta$-functions, and then search for
their common zeroes.  This computation is essentially nonperturbative,
because the resummation uses nonperturbative information on their
large-order behavior.  The comparison of the analyses of the MZM and
3D-$\overline{\rm MS}$ expansions provides nontrivial crosschecks of
the results. Details of these calculations can be found in
Refs.~\cite{BPV-05,BPV-03} (see also \cite{CP-04}).  In the following
we summarize the main results.

For $N=1$, the case relevant to $N_f=1$ QCD, ${\cal L}_{{\rm
U}(1)}$ reduces to the O(2) symmetric $\Phi^4$ theory, corresponding to
the XY universality class of superfluid transition in $^4$He, see, e.g.,
\cite{CHPRV-01}.  The determinant term related to the
axial anomaly, cf. Eq.~(\ref{LSUN}), plays the role of an external field, 
thus no continuous transition is expected, but a crossover.

The case $N=2$ without anomaly, 
cf. Eq.~(\ref{LUN}), relevant for $N_f=2$ QCD,
was originally analyzed by Pisarski and Wilczek
\cite{PW-84} within the $\epsilon\equiv 4-D$ expansion to one-loop
order, which means close to 4D.  No stable FP is found close to 4D, as
also in the case of symmetric matrix field relevant for $N_f=1$ aQCD.
Thus a naive extrapolation to 3D indicates first-order transitions.
However, in some physically interesting cases the extrapolation of
$\epsilon$-expansion calculations to $\epsilon=1$ fails to provide the
correct physical picture in 3D: for example, in the Ginzburg-Landau
model of superconductors \cite{superc}, and in O(2)$\otimes$O($N$)
theories of some frustrated spin models \cite{CPPV-04}.  Actually,
this happens also in the case of the LGW $\Phi^4$ theory (\ref{LUN})
with $N=2$.  Indeed, the analysis of the high-order series shows the
presence of stable FPs in both MZM and 3D-$\overline{\rm MS}$ schemes,
contradicting earlier analyses based on the $\epsilon$ expansion
around 4D.\footnote{This result was overlooked in
Ref.~\cite{BPV-03}. Some details can be found in Ref.~\cite{BPV-05}.}
These results imply the existence of 3D universality classes with SBP
U(2)$_L\otimes$U(2)$_R$$\rightarrow$U(2)$_V$ and
U(2)$\rightarrow$O(2),\footnote{The U(2)/O(2) universality class is
also the one of the normal-to-planar superfluid transition in $^3$He
\cite{DPV-04,BPV-05}.}  corresponding respectively to $N_f=2$ QCD and
$N_f=1$ aQCD in the case of suppressed U(1)$_A$ anomaly.

\begin{figure}
{\epsfig{width=15.2truecm,file=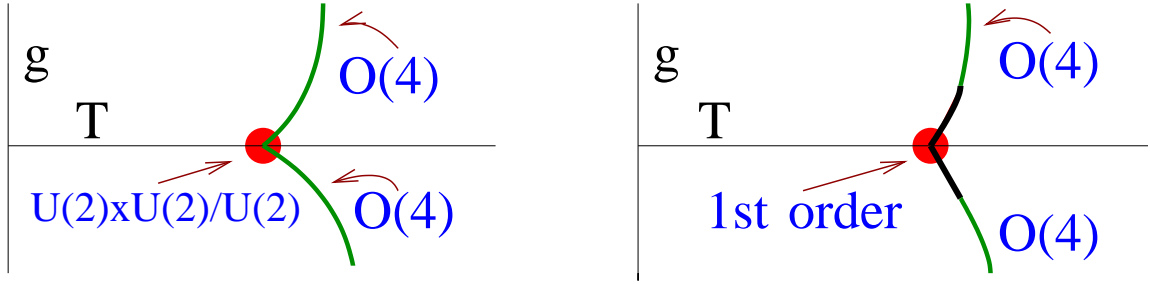}}
\caption{Possible phase diagrams in the $T$-$g$ plane 
for the LGW theory (\protect\ref{LSU2}) describing the transition of $N_f=2$ QCD,
in the case the transition at the multicritical point, i.e. for $g=0$,
is continuous (left) or first order (right).  Thick black lines
indicate first-order transitions. At their end points, thus for particular
values of $g$, the transition should be of mean-field type (apart from logarithms).
}
\label{pd}
\end{figure}

In two-flavor QCD, taking into account the U(1)$_A$ anomaly, SBP
 (\ref{qcdsb}) becomes equivalent to the one of the O(4) vector
universality class, i.e. SO(4)$\rightarrow$SO(3).  SBP
(\ref{aqcdsb}) of $N_f=1$ aQCD is instead equivalent to the one of the
O(3) vector universality class.  This means that, if the transition is
continuous, it must show the O(4) scaling behavior in $N_f=2$ QCD
and the O(3) one in $N_f=1$ aQCD. See, e.g.,
Ref.~\cite{PV-r} for a recent review on the O($N$) universality
classes.  Actually, the LGW $\Phi^4$ theory corresponding to these
cases is
\begin{eqnarray}
&&{\cal L}_{{\rm SU}(2)} ={\rm Tr} (\partial_\mu \Phi^\dagger)
(\partial_\mu \Phi) +r {\rm Tr} \Phi^\dagger \Phi + {u_0\over 4}
\left( {\rm Tr} \Phi^\dagger \Phi \right)^2 + {v_0\over 4} {\rm Tr}
\left( \Phi^\dagger \Phi \right)^2
\label{LSU2}\\
&&
+w_0 \left( {\rm det} \Phi^\dagger + {\rm det} \Phi \right)
+ {x_0\over 4} \left( {\rm Tr} \Phi^\dagger \Phi \right) 
         \left( {\rm det} \Phi^\dagger + {\rm det} \Phi \right) 
+ {y_0\over 4} \left[ ({\rm det} \Phi^\dagger)^2 + ({\rm det} \Phi)^2 \right] ,
\nonumber
\end{eqnarray}
where $w_0,x_0,y_0\sim g$ and $g$ parametrizes the effective breaking
of the U(1)$_A$ symmetry. If the anomaly is suppressed ($g=0$), then
$w_0=x_0=y_0=0$.  ${\cal L}_{{\rm SU}(2)}$ contains two quadratic
(mass) terms, therefore it describes several transition lines in the
$T$-$g$ plane, which meet at a multicritical point for $g=0$.  In the
case of QCD the multicritical behavior is controlled by the
U(2)$_L$$\otimes$U(2)$_R$ symmetric theory.  Possible phase diagrams
in the $T$-$g$ plane are shown in Fig.~\ref{pd}.  When $g\neq 0$ the
transition may be first order or continuous in the O(4) universality
class. Actually, we may also have a mean-field behavior
(apart from logarithms) for particular values of $g$, see Fig.~\ref{pd}.  
A similar scenario applies also to $N_f=1$ aQCD.

No stable FPs are found for $N>2$ in the LGW theory (\ref{LUN}).
Thus, neglecting the anomaly, transitions are always first order when
$N_f>2$ for QCD and $N_f>1$ for aQCD.  In most cases this result does
not change if we take into account the axial anomaly,
cf. Eq.~(\ref{LSUN}). The only exception is the case related to the
$N_f=2$ aQCD, where we find a stable FP corresponding to a 3D
SU(4)/SO(4) universality class, with critical exponents $\nu\approx
1.1$ and $\eta\approx 0.2$.  Note that, although SU(4)$\simeq$O(6),
the SBP of $N_f=2$ aQCD, i.e. SU(4)$\rightarrow$SO(4), differs from
that of the O(6) vector model, i.e. SO(6)$\rightarrow$ SO(5).  Thus,
according to the standard paradigm that relates the universality class
to the SBP, the corresponding universality classes and critical
behaviors must be different.

\begin{table}
\begin{center}
\begin{tabular}{ccc}
\hline\hline
\multicolumn{1}{c}{}& 
\multicolumn{1}{c}{U(1)$_A$ anomaly}& 
\multicolumn{1}{c}{suppressed anomaly at $T_c$}\\
\hline
\multicolumn{1}{c}{QCD}& 
\multicolumn{1}{c}{${\rm SU}(N_f)_L\otimes{\rm SU}(N_f)_R\rightarrow{\rm SU}(N_f)_V$}& 
\multicolumn{1}{c}{${\rm U}(N_f)_L\otimes{\rm U}(N_f)_R\rightarrow{\rm U}(N_f)_V$}\\ 
\hline
$N_f=1$ & crossover or first order & O(2) or first order  \\
$N_f=2$ & O(4) or first order & 
U(2)$_L\otimes$U(2)$_R$/U(2)$_V$  or first order  \\
$N_f\ge 3$ & first order & first order \\
\hline\hline
\multicolumn{1}{c}{aQCD}& 
\multicolumn{1}{c}{${\rm SU}(2N_f)\rightarrow{\rm SO}(2N_f)$}& 
\multicolumn{1}{c}{${\rm U}(2N_f)\rightarrow{\rm O}(2N_f)$}\\ 
\hline 
$N_f=1$ & O(3) or first order & U(2)/O(2) or first order  \\
$N_f=2$ & SU(4)/SO(4) or first order  & first order \\
\hline\hline
\end{tabular}
\caption{Summary of the RG predictions.
For each case we report the possible types of transition,
indicating the universality class when the transition can also be
continuous.
}
\label{summary}
\end{center}
\end{table}

The predictions of our RG analysis for the finite-$T$ chiral
transitions in QCD and aQCD are summarized in
Table~\ref{summary}. These transitions have also been investigated by lattice MC
simulations.  Overall, MC results for two-flavor QCD seem to favor a
continuous transition in the continuum limit.  However, a satisfactory
check of the O(4) scaling behavior has not been achieved yet.  Results
obtained using Wilson fermions appear consistent with a continuous
transition in the O(4) universality class
\cite{CP-PACS-01};  MC simulations using staggered fermions appear
more problematic \cite{KLP-01,KS-01,EHMS-01,pi12-05,EHS-05}.  
Unlike $N_f=2$ QCD, the transition scenario  appears settled 
for $N_f\ge 3$: MC simulations \cite{KLP-01,IKSY-96} show first-order
transitions, in agreement with the RG predictions. Finally, in the
case of $N_f=2$ aQCD the available MC results \cite{KL-99,EHS-05}
favor a continuous transitions. But they are not yet sufficiently
precise to check the critical behavior of the 3D SU(4)/SO(4)
universality class.


\begin{thebibliography}{99}


\bibitem{reviews}
See, e.g., 
F. Wilczek, \emph{QCD in extreme conditions}, {\tt hep-ph/0003183};
F. Karsch,  \emph{Lectures on Quark Matter}, {\tt hep-lat/0106019}.

\bibitem{KL-99}
F. Karsch and M. L\"utgemeier, \emph{Deconfinement and chiral symmetry
restoration in an SU(3) gauge theory with adjoint fermions},
\emph{Nucl. Phys.} {\bf B550} (1999) 449 [{\tt hep-lat/9812023}].

\bibitem{EHS-05}
 J. Engels, S. Holtmann, and T. Schulze,
\emph{Scaling and Goldstone effects in a QCD with two flavours of adjoint quarks},
\emph{Nucl. Phys.} {\bf B724} (2005) 357 
[{\tt hep-lat/0505008}]; 
\emph{The chiral transition of $N_f=2$ QCD with fundamental and adjoint
fermions}, \emph{PoS (LAT2005)} 148
[{\tt hep-lat/0509010}].

\bibitem{PW-84}
R.D. Pisarski and F. Wilczek, \emph{Remarks on the chiral phase
transition in chromodynamics},
\emph{Phys. Rev.} {\bf D29} (1984) 338.


\bibitem{BPV-05}
F. Basile, A. Pelissetto, and E. Vicari, \emph{The finite-temperature
chiral transition in QCD with adjoint fermions},
\emph{JHEP} {\bf 02} (2005) 044 [{\tt hep-th/041202}].

\bibitem{BPV-03}
A. Butti, A. Pelissetto, and E. Vicari, \emph{On the nature of the
finite-temperature transition in QCD},
\emph{JHEP} {\bf 08} (2003) 029 
[{\tt hep-ph/0307036}].


\bibitem{CP-04}
P. Calabrese and P. Parruccini,
\emph{Five-loop $\epsilon$ expansion for $U(n)\otimes U(m)$ models: 
finite-temperature phase transition in light QCD},
\emph{JHEP} {\bf 05} (2004) 018 
[{\tt hep-ph/0403140}].

\bibitem{CHPRV-01} 
M.~Campostrini, M.~Hasenbusch, A.~Pelissetto, P.~Rossi, and E.~Vicari, 
\emph{Critical behavior of the three-dimensional $XY$ universality class},
\emph{Phys. Rev.}   {\bf B 63} (2001) 214503 [{\tt cond-mat/0010360}].
%%CITATION = PHRVA,B63,214503;%%

\bibitem{PV-r} 
A. Pelissetto and E. Vicari,
\emph{Critical phenomena and renormalization-group theory},
\emph{Phys. Rep.} {\bf 368} (2002) 549 [{\tt cond-mat/0012164}].
%%CITATION = COND-MAT 0012164;%%

\bibitem{superc}
S. Mo, J. Hove, and A. Sudb\o,
\emph{Order of the metal-to-superconductor transition},
\emph{Phys. Rev.}   {\bf B65} (2002) 104501 [{\tt cond-mat/0109260}].

\bibitem{CPPV-04}
P. Calabrese, P. Parruccini, A. Pelissetto, and E. Vicari, 
\emph{Critical behavior of O(2)$\otimes$O($N$) symmetric models},
\emph{Phys. Rev.} {\bf B70} (2004) 174439 [{\tt cond-mat/0405667}].
%%CITATION = COND-MAT 0405667;%

\bibitem{DPV-04}
M. De Prato, A. Pelissetto, and E. Vicari, 
\emph{Normal-to-planar superfluid transition in ${}^3$He},
\emph{Phys. Rev.} {\bf B70} (2004) 214519  [{\tt cond-mat/0312362}].
%%CITATION = COND-MAT 0312362;%%

\bibitem{CP-PACS-01}
A. Ali Khan \emph{et al}. (CP-PACS Collaboration), 
\emph{Phase structure and critical temperature of two-flavor QCD 
with renormalization-group improved gauge action and clover 
improved Wilson action},
\emph{Phys. Rev.} {\bf D63} (2001) 034502  [{\tt hep-lat/0008011}].


\bibitem{KLP-01}
F. Karsch, E. Laermann, and A. Peikert,
\emph{Quark mass and flavor dependence of the QCD phase transition},
\emph{Nucl. Phys.} {\bf B605} (2001) 579 [{\tt hep-lat/0012023}].

\bibitem{KS-01}
J. B. Kogut and D. K. Sinclair,
\emph{Scaling behavior at the $N_t=6$ chiral phase transition
for 2-flavor lattice QCD with massless staggered quarks and an irrelevant 4-fermion
interaction},
\emph{Phys. Rev.} {\bf D64} (2001) 034508 [{\tt hep-lat/0104011}].

\bibitem{EHMS-01}
J. Engels, S. Holtmann, T. Mendes, and T. Schulze,
\emph{Finite-size-scaling behavior for $3d$ O(4) and O(2)
spin models and QCD},
\emph{Phys. Lett.} {\bf B514} (2001) 299 [{\tt hep-lat/0105028}].


\bibitem{pi12-05}
 M. D'Elia, A. Di Giacomo, and C. Pica,
\emph{Two flavor QCD and Confinement},
{\tt hep-lat/0503030}.


\bibitem{IKSY-96}
Y. Iwasaki, K. Kanaya, S. Sakai, and T. Yoshi\'e,
\emph{Chiral phase transition in lattice QCD with Wilson quarks},
\emph{Z. Physik} {\bf C71} (1996) 337 [{\tt hep-lat/9504019}].




\end{thebibliography}
\end{document}